\documentclass[12pt]{article}
\usepackage{amsfonts,amsthm,amsmath,amssymb,upgreek,bm}
\usepackage{hyperref}
\usepackage[paper=letterpaper,margin=.85in]{geometry}
\usepackage{graphicx}
\usepackage{units}
\usepackage{float}
\usepackage[export]{adjustbox}
\usepackage{xcolor}
\usepackage[shortlabels]{enumitem}
\usepackage[mathscr]{euscript}
\usepackage[normalem]{ulem}
\usepackage{bm}
\usepackage{tikz}
\usepackage{tikzsymbols}
\usepackage{pgfplots}
\usepackage{ytableau}
\usepackage{slashed}
\usepackage{mathtools}
\usetikzlibrary{decorations.pathmorphing}
\usetikzlibrary{arrows.meta}

\newcommand{\be}{\begin{equation}}
\newcommand{\ee}{\end{equation}}
\newcommand{\bea}{\begin{eqnarray}}
\newcommand{\eea}{\end{eqnarray}}

\renewcommand{\i}{\mathrm{i}}
\renewcommand{\d}{\mathrm{d}}

\numberwithin{equation}{section}

\def\tr{\text{Tr}}
\def\Tr{\text{Tr}}

\def\CP {\mathbb{C P}}
\def\bC {\mathbb{C}}

\newcommand{\pa}{\partial}

\def\be{\begin{equation}}
\def\ee{\end{equation}}
\def\bea{\begin{eqnarray}}
\def\eea{\end{eqnarray}}
\def\ie{\begin{equation}\begin{aligned}}
\def\fe{\end{aligned}\end{equation}}

\newcommand{\cH}{{\mathcal H}}

\newcommand{\cO}{{\mathcal O}}

\begin{document}
\thispagestyle{empty}

\vspace*{2.5cm}
\begin{center}

{\bf {\LARGE Bulk thimbles dual to trace relations}}

\begin{center}

\vspace{1cm}

{\bf Ji Hoon Lee$^1$ and Douglas Stanford$^2$}\\
 \bigskip \rm

\bigskip 

${}^1$Institut f\"{u}r Theoretische Physik, ETH Zurich,\\ CH-8093 Z\"urich, Switzerland

\bigskip

${}^2$Stanford Institute for Theoretical Physics,\\Stanford University, Stanford, CA 94305

\rm
  \end{center}

\vspace{2.5cm}
{\bf Abstract}
\end{center}
\begin{quotation}
\noindent

The maximal giant graviton is a D-brane wrapping a maximal $S^3\subset S^5$ within $\text{AdS}_5\times S^5$. It represents an upper bound on the $R$ charge that can be carried by certain bulk states. We study the maximal giant and its half-BPS fluctuations, motivated by a recent proposal \cite{Lee:2023iil} connecting these fluctuations to trace relations in the boundary theory. In a computation of the partition function of half-BPS states, we find that the maximal giant is an unstable saddle point and that its Lefschetz thimble corresponds to the quantization of an imaginary phase space. The states resulting from the quantization of this phase space contribute negatively to the partition function and can be regarded as bulk duals of trace relations. Finally, we study a model for a path integral that would connect together components of the bulk half-BPS field space with different numbers of giants.

\end{quotation}

\setcounter{page}{0}
\setcounter{tocdepth}{2}
\setcounter{footnote}{0}

\newpage

\parskip 0.1in
 
\setcounter{page}{2}
\tableofcontents

\section{Introduction}
The half-BPS sector of the $U(N)$ $\mathcal{N} = 4$ SYM theory is spanned by multi-traces $\tr(Z^{k_1})\tr(Z^{k_2})\dots$ where $Z$ is a scalar that carries charge one under a particular $R$ charge. The combinations of multi-traces that diagonalize the inner product are the so-called ``Schur polynomials'' \cite{Corley:2001zk}.\footnote{A nontrivial special case corresponding to subdeterminants was worked out earlier in \cite{Balasubramanian:2001nh}.} These can be labeled by Young diagrams with $R$ total boxes, for example
\be
\begin{tikzpicture}[scale=0.25, rotate=0, baseline={([yshift=-0.15cm]current bounding box.center)}]
  \draw[thick] (0,0) -- (1,0) -- (1,-1) -- (0,-1) -- (0,0);
\end{tikzpicture}
\leftrightarrow \tr Z \hspace{30pt} 
\begin{tikzpicture}[scale=0.25, rotate=0, baseline={([yshift=-0.15cm]current bounding box.center)}]
  \draw[thick] (0,0) -- (1,0) -- (1,-2) -- (0,-2) -- (0,0);
  \draw[thick] (0,-1) -- (1,-1);
\end{tikzpicture}
\leftrightarrow \tr (Z^2) - (\tr Z)^2
\hspace{30pt}
\begin{tikzpicture}[scale=0.25, rotate=90, baseline={([yshift=-0.125cm]current bounding box.center)}]
  \draw[thick] (0,0) -- (1,0) -- (1,-2) -- (0,-2) -- (0,0);
  \draw[thick] (0,-1) -- (1,-1);
\end{tikzpicture}
\leftrightarrow \tr (Z^2) + (\tr Z)^2.
\ee
Any Young diagram with a column that is taller than $N$ boxes corresponds to a {\it trace relation} operator that is zero due to identities relating different traces of a finite rank matrix. For example, the operator $\tr (Z^2) - (\tr Z)^2$ is zero if $N = 1$.

The partition function of half BPS states is a sum over such Young diagrams, and it can be organized into a series known as the giant graviton expansion \cite{Gaiotto:2021xce,Imamura:2021ytr} (see also \cite{Arai:2019xmp,Arai:2020qaj,Lee:2022vig}), where the $N = \infty$ answer is corrected by a series of terms suppressed by powers of $q^N$.\footnote{\cite{Murthy:2022ien} introduced a distinct \cite{Liu:2022olj} but related \cite{Eniceicu:2023uvd} expansion that was related to eigenvalue instantons in \cite{Chen:2024cvf}.} A baby example of this is contained in the sum over single-column Young diagrams:
\begin{align}\label{finite}
\sum_{\text{single column states}} q^R &= 1 + q + \dots + q^{N} \\ &= \frac{1 }{1-q} - \frac{q^{N+1}}{1-q}.\label{finite2}
\end{align}

In the bulk theory, the multi-traces that correspond to the terms in this sum are superpositions of single and multi-graviton states. As the $R$ charge grows larger, the superposition involves a large number of non-orthogonal terms and the graviton description stops being useful. However, for certain classes of Young diagrams, there are still simple bulk descriptions. In particular, a single tall column corresponds to a {\it giant graviton} \cite{McGreevy:2000cw,Grisaru:2000zn}, which means a brane configuration with a D3 brane wrapped on an $S^3$ within the $S^5$ of $AdS_5\times S^5$.\footnote{Long rows are dual giant gravitons \cite{Grisaru:2000zn}, where the D3 brane wraps an $S^3$ within $AdS_5$. Young diagrams without any nearby corners are smooth LLM geometries formed from a coherent state of gravitons \cite{Lin:2004nb}.} The D-brane explanation for the upper limit on the sum (\ref{finite}) is that there is a maximal $S^3$ within $S^5$ that the D3 brane can wrap. In this paper we will clarify the D-brane explanation for the negative term in (\ref{finite2}).

From the boundary perspective, the second line (\ref{finite2}) organizes the answer into the $N = \infty$ answer minus the effect of trace relations:
\be
\underbrace{1 + q + \dots + q^N}_{\text{physical states}} = \underbrace{\frac{1 }{1-q}}_{\text{$N=\infty$ answer}} \hspace{10pt} - \hspace{10pt}\underbrace{\frac{q^{N+1}}{1-q}}_{\text{trace relations}}
\ee
From the bulk perspective, we will see that the negative term arises as the contribution of the maximal giant graviton. The main subtlety that allows it to contribute a negative answer is that the maximal giant graviton is an unstable saddle point of the functional integral that computes the partition function of half-BPS states:
\be\label{introbeta}
Z_{\text{BPS}}(q) = \lim_{\beta\to\infty}\text{Tr}(e^{-\beta(H-R)}q^R).
\ee
The Lefschetz thimble of this unstable saddle is the path-integral quantization of the phase space of the giant graviton, with two coordinates Wick-rotated (continued to imaginary values). The result is the negative term
\be
-\frac{q^{N+1}}{1-q} = -q^{N+1} - q^{N+2}+\dots
\ee
Terms on the RHS can be identified with specific states in the quantization of the Wick-rotated phase space, and we interpret them as bulk duals of trace relations.

Perturbative excitations of the maximal giant have been studied recently in \cite{Lee:2023iil,Eleftheriou:2023jxr,Beccaria:2024vfx,Gautason:2024nru} and we will use some of their results. In section \ref{classical} we study the relevant fluctuations of the maximal giant graviton in a classical approximation. In section \ref{quantum} we treat them quantum mechanically. In section \ref{discussion} we study a model for a path integral that would connect together components of the bulk half-BPS field space with different numbers of giants.

\section{Classical analysis}\label{classical}
\subsubsection*{Ansatz and Lagrangian}
Let's start by reviewing the Lagrangian for the giant graviton \cite{McGreevy:2000cw}. The $AdS_5\times S^5$ background is
\begin{align}
\d s^2 &= \Big[-\cosh^2(\rho) \d t^2 + \d\rho^2 + \sinh^2(\rho) \d\widehat\Omega_3^2\Big] + \Big[\cos^2(\psi) \d\phi^2  + \d\psi^2+ \sin^2(\psi)\d\Omega_3^2\Big]\\
C_4 &= \sinh^4(\rho) \d t\wedge \d\widehat\Omega_3 + \sin^4(\psi) \d\phi\wedge \d\Omega_3.
\end{align}
The group of $R$-symmetries corresponds in the bulk theory to isometries on the $S^5$. We will focus on a particular $R$ charge that generates translations in $\phi$. 

The scalar part of the worldvolume action of a D3 brane is
\be
S = - \frac{N}{2\pi^2}\int_{\text{D3}} \Big(\d^4 x \sqrt{-g_{\text{D3}}} - C_4\Big).
\ee
To study the half-BPS solutions, we can make an ansatz \cite{McGreevy:2000cw,Grisaru:2000zn} in which the brane sits at at the center of $AdS$ with $\rho = 0$ and wraps $\Omega_3\supset S^5$ homogeneously. The brane is localized in the remaining coordinates $\phi,\psi$ on the $S^5$, which we allow to depend on time. The Lagrangian for $\phi(t)$ and $\psi(t)$ is
\begin{align}
\mathcal{L} &= N\left\{-\sin^3(\psi)\sqrt{1 - \cos^2(\psi)\dot\phi^2 - \dot\psi^2} + \sin^4(\psi)\dot\phi \right\}.
\end{align}
The coordinate $\psi$ runs from $0$, where our D3 brane would be pointlike to $\pi/2$, where it would be a {\it maximal giant graviton}, wrapping a maximal $S^3$. 

This ansatz is capable of describing a small class of very symmetric states of the giant graviton. The reason we are limiting ourselves to this class is that it contains all of the half BPS states. All of the other fluctuations were considered \cite{Gautason:2024nru} (see table 2) but they have a gap in $H-R$ so for large $\beta$ in (\ref{introbeta}) they are projected into their ground state.

\subsubsection*{Approximating near the maximal giant}
We will focus on the region near the maximal giant, zooming in by writing \cite{Lee:2023iil,Eleftheriou:2023jxr}
\be
\psi = \frac{\pi}{2} - \frac{r}{\sqrt{N}}.
\ee
To quadratic order in $r$, the Lagrangian is
\be\label{Lagrangian}
\mathcal{L} = N\left(-1 + \dot\phi\right) +\frac{r^2\dot\phi^2 + \dot r^2}{2} + \frac{3}{2}r^2 - 2 r^2\dot\phi.
\ee
In this approximation, the $R$ charge and Hamiltonian are
\begin{align}
R &= \frac{\partial\mathcal{L}}{\partial\dot\phi} \\ &= N - 2r^2 + r^2\dot\phi,\label{Rcharge}
\end{align}
and
\begin{align}
H &= \dot\phi R + \dot{r} p  - \mathcal{L} \\ &= 2R - N + \frac{p^2+r^2}{2} + \frac{(R-N)^2}{2r^2},\label{D3H}
\end{align}
where $p = \dot r$ is the momentum conjugate to $r$. 

\subsubsection*{Reducing to the BPS phase space}
We have a four-dimensional phase space with symplectic form
\be
\omega = \d p \wedge \d r + \d R \wedge \d \phi.
\ee
To reduce to the BPS submanifold of this phase space, we impose $H = R$, which gives
\be
p^2 = -\frac{(R-N+r^2)^2}{r^2} \hspace{20pt} \implies \hspace{20pt} p = 0 \text{ and } R = N-r^2.\label{BPScond}
\ee
So the single equation $H = R$ effectively imposes two conditions, leaving a two-dimensional BPS phase space, parametrized by $(\phi,R)$ or $(\phi,r)$ with the understanding that $R = N-r^2$. We will also find it useful to parametrize using Cartesian-type coordinates
\be
r e^{\i\phi} = x_1 + \i x_2.
\ee
The symplectic form in the various coordinates is
\begin{align}
\omega = \d R \wedge \d\phi  = 2r\d r\wedge\d\phi  = 2 \d x_1\wedge \d x_2,
\end{align}
and the $R$ charge is
\be
R = N - r^2 = N - x_1^2-x_2^2.
\ee

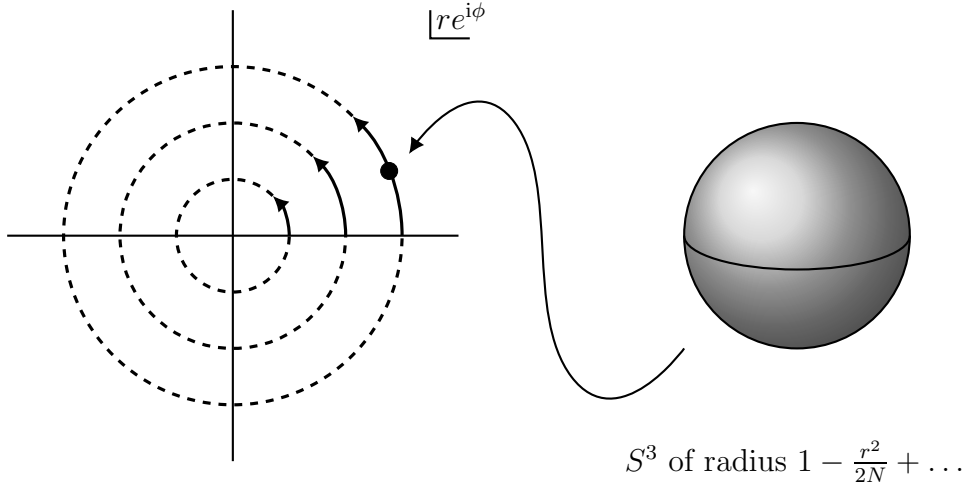
\begin{figure}
\begin{center}
\begin{tikzpicture}[scale=0.75, rotate=0, baseline={([yshift=-0.15cm]current bounding box.center)}]
  \draw[thick] (-4,0) -- (4,0);
  \draw[thick] (0,-4) -- (0,4);
  \draw[thick] (3.5,4) -- (3.5,3.5) -- (4.2,3.5);
  \node[] at (4.05,3.9) {$re^{\i\phi}$};
  \draw[very thick,->,>={Latex[length=6pt, width=6pt]}] (2,0) arc(0:45:2);
  \draw[very thick, dashed] (1.414,1.414) arc(45:360:2);
  \draw[very thick,->,>={Latex[length=6pt, width=6pt]}] (1,0) arc(0:45:1);
  \draw[very thick, dashed] (.5*1.414,.5*1.414) arc(45:360:1);
  \draw[very thick,->,>={Latex[length=6pt, width=6pt]}] (3,0) arc(0:45:3);
  \draw[very thick, dashed] (1.5*1.414,1.5*1.414) arc(45:360:3);
  \draw[fill] (2.771,1.148) circle(.15);
  \shade[ball color = gray!40, opacity = 0.4] (10,0) circle (2cm);
  \draw[thick] (10,0) circle (2cm);
  \draw[thick] (8,0) arc (180:360:2 and 0.6);
  \node[] at (10,-4) {$S^3 \text{ of radius }1 - \frac{r^2}{2N} + \dots$};
  \draw[thick, ->,>={Latex[length=6pt, width=6pt]}] plot [smooth, tension=1] coordinates { (8,-2) (6,-2.5) (5,2) (2.771+.35,1.148+.3)};
\end{tikzpicture}
\end{center}
\caption{The classical solutions corresponding to the BPS states of the giant graviton. On the left we show the two dimensional plane of $re^{\i\phi} = x_1 + \i x_2$ -- this corresponds to a local description of two of the dimensions of the $S^5$. The other three dimensions of the $S^5$ consist of an $S^3$ of radius $\approx 1 - r^2/(2N)$. The D3 brane wraps this $S^3$ and sits at a point in the $re^{\i\phi}$ plane. As a function of Lorentzian time, this point orbits at a constant rate around the origin. From (\ref{Rcharge}) and (\ref{BPScond}) the angular velocity is $\dot\phi = 1$.}\label{fig1}
\end{figure}

\subsubsection*{BPS partition function in the classical approximation}
We can integrate over the BPS phase space to compute the partition function
\begin{align}\label{lineone}
Z^{\text{classical}}_{\text{BPS}}(q) = \int \frac{\d\phi\d R}{2\pi} q^R &= q^N \int_0^?\d(r^2)q^{-r^2}.
\end{align}
Note that the integral diverges (for the case of interest $|q|<1$). This divergence is because we have approximated near the maximal giant $r = 0$ -- in a more complete description the large $r$ behavior would be modified once $r^2\sim N$. We won't worry about this because our interest is in the contribution associated to the maximal giant itself. We claim that this is simply
\be
Z^{\text{classical}}_{\text{BPS}}(q)\supset -\frac{q^N}{\log\frac{1}{q}}.
\ee
This can be seen in different ways. In polar coordinates $(\phi,r)$ or $(\phi,R)$ the above is the contribution from the endpoint at $r = 0$ or $R = N$, and the minus sign is because it is the contribution from a lower endpoint. In smooth Cartesian coordinates $re^{\i\phi} = x_1 + \i x_2$ there is no endpoint but the maximal giant $x_1 = x_2 = 0$ is an unstable saddle point. The Lefshetz thimble associated to this saddle can be described by Wick-rotating both $x_1,x_2$ integrals to imaginary values. This thimble contributes with coefficient one (see appendix \ref{app:unstable}), and the integral (\ref{lineone}) becomes
\begin{align}\label{classicalans}
Z^{\text{classical}}_{\text{BPS}}(q) &= [\text{other thimbles}] + q^N\int_{-\i\infty}^{\i\infty}\int_{-\i\infty}^{\i\infty} \frac{2\d x_1\d x_2}{2\pi}q^{-(x_1^2 + x_2^2)} \\ &=[\text{other thimbles}]  - \frac{q^N}{\log\frac{1}{q}}.
\end{align}
The minus sign comes from Wick-rotating two coordinates, $\i^2 = (-1)$.

\section{Quantum analysis}\label{quantum}
We will now study the partition function of BPS states in the quantum theory, still working with the approximate theory with Lagrangian (\ref{Lagrangian}) and Hamiltonian (\ref{D3H}).

\subsubsection*{Mapping to two harmonic oscillators}
To quantize the Hamiltonian (\ref{D3H}) with as little work as possible, it is helpful to realize that it is closely related to a pair of harmonic oscillators written in polar coordinates. Consider two harmonic oscillators $x_1,x_2$ with angular momentum $\ell$. In polar coordinates, their Hamiltonian (we set the frequency $\omega = 1$) is 
\be
h = \frac{p^2 + r^2}{2} + \frac{\ell^2}{2r^2}
\ee
where $r$ is the radial coordinate $r^2 = x_1^2 + x_2^2$ and $p = p_r$ is its conjugate momentum. Our Hamiltonian (\ref{D3H}) can be mapped to this one by writing
\begin{align}\label{transofHandR}
H &= h + 2\ell + N\\
R &= \ell + N,
\end{align}
where, in Cartesian coordinates,
\begin{align}
h &= \frac{p_1^2 + x_1^2}{2} + \frac{p_2^2 + x_2^2}{2}\\
\ell &= x_1 p_2 - x_2 p_1.
\end{align}

One can go further and decouple the two harmonic oscillators by defining
\be
X_1 = \frac{x_1 - p_2}{\sqrt{2}} \hspace{20pt} X_2 = \frac{x_2-p_1}{\sqrt{2}}, \hspace{20pt} P_1 = \frac{x_2 + p_1}{\sqrt{2}}, \hspace{20pt} P_2 = \frac{x_1+p_2}{\sqrt{2}}.
\ee
The $(X_1,P_1)$ and $(X_2,P_2)$ variables are canonically conjugate, and in terms of these variables
\begin{align}\label{Hform}
H - R &= X_2^2 + P_2^2\\
R &= N + \frac{X_2^2+P_2^2}{2}- \frac{X_1^2 + P_1^2}{2}.\label{Rform}
\end{align}
The spectra of $H$ as a function of $R$ and $H-R$ as a function of $R$ are plotted below.
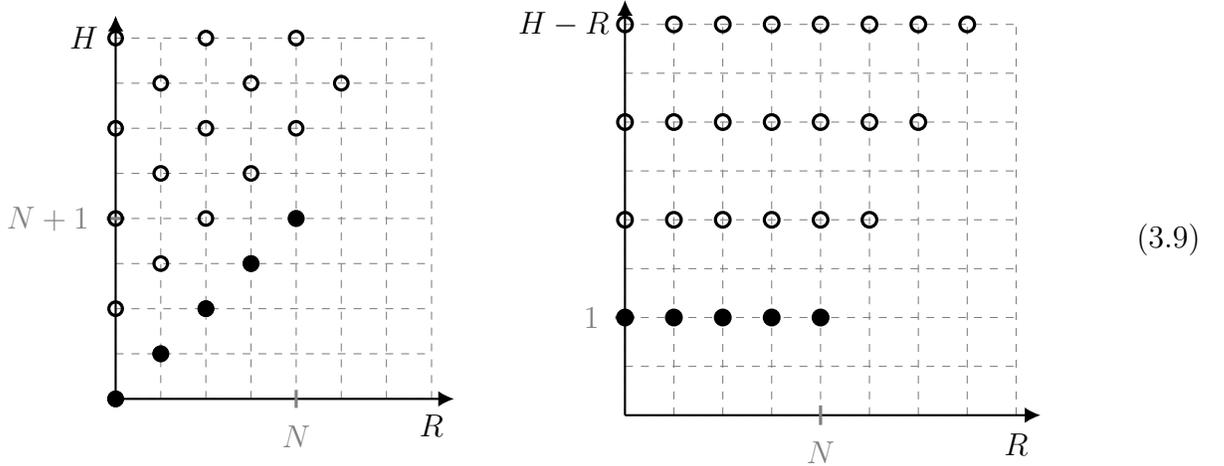
\be
\begin{tikzpicture}[scale=0.6, rotate=0, baseline={([yshift=-0.15cm]current bounding box.center)}]
\newcommand*{\xMin}{-3}
\newcommand*{\xMax}{4}
\newcommand*{\yMin}{-3}
\newcommand*{\yMax}{5}
  \foreach \i in {\xMin,...,\xMax} {
        \draw [gray, dashed] (\i,\yMin) -- (\i,\yMax) ;
    }
  \foreach \i in {\yMin,...,\yMax} {
        \draw [gray, dashed] (\xMin,\i) -- (\xMax,\i);
    }
  \draw[thick, ->,>={Latex[length=6pt, width=6pt]}] (\xMin,\yMin) -- (\xMax+.5,\yMin);
  \draw[thick, ->,>={Latex[length=6pt, width=6pt]}] (\xMin,\yMin) -- (\xMin,\yMax+.5);
  \draw[very thick, gray] (\xMin-.2,1) -- (\xMin+.2,1) node[left = 9] {$N+1$};
  \draw[very thick, gray] (\xMin/2+\xMax/2+.5,\yMin-.2) -- (\xMin/2+\xMax/2+.5,\yMin+.2) node[below = 9] {$N$};
  \node[] at (\xMin -.7,.5*\yMin + .5*\yMax+4) {$H$};
  \node[] at (.5*\xMin + .5*\xMax +3.5,\yMin -.6) {$R$};
  \renewcommand*{\xMin}{-5}
  \renewcommand*{\yMin}{-5}
  \foreach \h in {1,...,3} {
      \foreach \r in {1,...,\h} {
        \draw[very thick] (\xMin + 2*\r,\yMin + 2*\h) circle(.15);
        }
    }
   \foreach \h in {1,...,2} {
      \foreach \r in {1,...,\h} {
        \draw[very thick] (\xMin + 2*\r+1,\yMin + 2*\h+1) circle(.15);
        }
    }
    \foreach \r in {1,...,3} {\draw[very thick] (\xMin + 2*\r,\yMin + 8) circle(.15); }
    \foreach \r in {1,...,2} {\draw[very thick] (\xMin + 2*\r+1,\yMin + 7) circle(.15); }
     \foreach \r in {1,...,3} {\draw[very thick] (\xMin + 2*\r+1,\yMin + 9) circle(.15); }
    \foreach \r in {1,...,3} {\draw[very thick] (\xMin + 2*\r,\yMin + 10) circle(.15); }
    \foreach \r in {2,...,6} {\draw[very thick, fill] (\xMin + \r,\yMin + \r) circle(.15); }
\end{tikzpicture}
\hspace{20pt}
\begin{tikzpicture}[scale=0.65, rotate=0, baseline={([yshift=-0.15cm]current bounding box.center)}]
\newcommand*{\xMin}{-3}
\newcommand*{\xMax}{5}
\newcommand*{\yMin}{-1}
\newcommand*{\yMax}{7}
  \foreach \i in {\xMin,...,\xMax} {
        \draw [gray, dashed] (\i,\yMin) -- (\i,\yMax) ;
    }
  \foreach \i in {\yMin,...,\yMax} {
        \draw [gray, dashed] (\xMin,\i) -- (\xMax,\i);
    }
  \draw[thick, ->,>={Latex[length=6pt, width=6pt]}] (\xMin,\yMin) -- (\xMax+.5,\yMin);
  \draw[thick, ->,>={Latex[length=6pt, width=6pt]}] (\xMin,\yMin) -- (\xMin,\yMax+.5);
  \draw[very thick, gray] (\xMin-.2,1) -- (\xMin+.2,1) node[left = 9] {$1$};
  \draw[very thick, gray] (\xMin/2+\xMax/2,\yMin-.2) -- (\xMin/2+\xMax/2,\yMin+.2) node[below = 9] {$N$};
  \node[] at (\xMin -1.25,.5*\yMin + .5*\yMax+3+1) {$H-R$};
  \node[] at (.5*\xMin + .5*\xMax +4,\yMin -.6) {$R$};
  \renewcommand*{\xMin}{-5}
  \renewcommand*{\yMin}{0}
  \foreach \r in {2,...,6} {\draw[very thick, fill] (\xMin + \r,1) circle(.15); }
  \foreach \r in {2,...,7} {\draw[very thick] (\xMin + \r,3) circle(.15); }
  \foreach \r in {2,...,8} {\draw[very thick] (\xMin + \r,5) circle(.15); }
  \foreach \r in {2,...,9} {\draw[very thick] (\xMin + \r,7) circle(.15); }
\end{tikzpicture}
\ee
The grid size is one unit in both directions, and we have filled some of the circles to emphasize the states that will become the BPS states.\footnote{The wave functions for the BPS state (filled dots) are
\be\label{physicalStates}
\Psi(r,\phi) = r^{N-R} e^{\i R\phi} e^{-r^2/2},
\ee
and the constraint $R \le N$ is needed for normalizability. These correspond to the classical solutions with $\dot\phi = 1$. There are also classical solutions where $\dot\phi= 3$, $R > N$ \cite{Grisaru:2000zn} and $H = 1+3R-2N$. These are non-BPS states with wave functions
\be
\Psi(r,\phi) = r^{R-N} e^{\i R\phi}e^{-r^2/2}.
\ee} \footnote{Note that the BPS states have $H - R = 1$, rather than the expected BPS value $H - R = 0$. This would be corrected by including the zero point energies of all of the other modes, which will of course also lead to further excited states with larger values of $H-R$.} Note that $H$ is not bounded from below. This is because we have approximated the system near the maximal giant $r\sim 0$. This description is accurate for states with $H$ and $R$ close to $N$, but it breaks down for lower energies. However, this description {\it does} accurately give a sharp transition in behavior at $R = N$. In particular, the sum of $q^R$ over the BPS states gives
\be
\tr_{\text{BPS}}(q^R) = \dots + q^{N-2} + q^{N-1} + q^N.
\ee

The nice feature of the $X_1,X_2,P_1,P_2$ variables is that the BPS subspace is just the ground state of the $(X_2,P_2)$ oscillator tensored with any state of the $(X_1,P_1)$ oscillator. To clarify the interpretation of $X_1,P_1$, note that they are equal to $x_1,x_2$ after projection $\Pi$ onto the ground state of the $X_2,P_2$ oscillator:
\be\label{LLL}
(1\otimes \Pi) x_1 (1\otimes \Pi) = \frac{X_1}{\sqrt{2}}\otimes \Pi, \hspace{20pt} (1\otimes \Pi) x_2 (1\otimes \Pi) = \frac{P_1}{\sqrt{2}}\otimes \Pi.
\ee
This transmuation of a configuration space $(x_1,x_2)$ into a phase space $(X_1,P_1)$ is familiar from the Landau problem, and our two oscillators are actually equivalent to that problem \cite{Eleftheriou:2023jxr}, where the BPS states correspond to the lowest Landau level.

\subsubsection*{Quantum path integral}
Naively, one can project onto the BPS states by computing:
\begin{align}
\text{Tr}_{\text{BPS}}(q^R) &\stackrel{?}{=} \lim_{\beta\to\infty}\text{Tr}\left( e^{-\beta(H-R)}q^R\right).\label{limit}
\end{align}
This would be the right thing to do in the full theory, but our truncated theory of two harmonic oscillators, the D3 brane does not have the correct zero point energy: BPS states have $H-R = 1$ instead of $H-R = 0$.\footnote{One obtains the correct zero-point energy by including contributions to the Casimir energy from all fluctuation fields, as shown in \cite{Beccaria:2024vfx}. The analysis there in the refined Schur sector can be specialized to the half-BPS sector by redefining $u \to q^{-1} u$ and then projecting $q \to 0$. This projects the Hilbert space on to states with $H + J + \bar{J} - R_1 + R_2 = 0$, in addition to those that satisfy the $1/16$-BPS condition $H - 2 \bar{J} - R_1 - R_2 - R_3 = 0$. The half-BPS complex scalar $X$ with $R_1 = 1$ in $\mathcal{N}=4$ SYM is the only field that satisfies these two constraints.} So to compensate for this we should instead compute
\begin{align}
\text{Tr}_{\text{BPS}}(q^R) &= \lim_{\beta\to\infty}\text{Tr}\left( e^{-\beta(H-R-1)}q^R\right).
\end{align}
In the large $\beta$ limit, the $(X_2,P_2)$ oscillator will be projected into its ground state, and we will be left with a path integral for the $(X_1,P_1)$ oscillator. The phase space path integral formula is
\begin{align}
\text{Tr}_{\text{BPS}}(q^R) = q^N\int \mathcal{D}X_1\mathcal{D}P_1 \exp\left\{\int_0^1\d\tau \left[\i P_1 \dot X_1 + \frac{\log\frac{1}{q}}{2}(X_1^2 + P_1^2-1)\right]\right\}.\label{pathint7}
\end{align}

The classical analysis we did previously can be recovered by restricting to the constant modes of $X_1(\tau) \to \sqrt{2}x_1$ and $P_1(\tau) \to\sqrt{2}x_2$ (see (\ref{LLL})). After expanding $X_1$ and $P_1$ in Fourier modes, the full path integral (\ref{pathint7}) is
\begin{align}
\text{Tr}_{\text{BPS}}(q^R) &= q^{N+\frac{1}{2}}\prod_{n = -\infty}^\infty \int \frac{\d^2 z_n}{2\pi} e^{(\pi\i n + \frac{\log\frac{1}{q}}{2})|z_n|^2}, \hspace{20pt} z_n = (X_1)_n + \i (P_1)_n\\
&\supset q^{N+\frac{1}{2}}\prod_{n = -\infty}^\infty \frac{2}{-\log\frac{1}{q}-2\pi\i n}\\
&=q^{N+\frac{1}{2}} \cdot \frac{-2}{\log\frac{1}{q}}\cdot\prod_{n = 1}^\infty \frac{4}{\log^2\frac{1}{q}+(2\pi n)^2}\\
&\to -\frac{q^{N+1}}{1-q}.\label{toExpand}
\end{align} 
The $``\supset"$ in the second line means that we are only showing the contribution from the thimble associated to the maximal giant saddle point. In the third line, we combined together the $n < 0$ and $n > 0$ terms. After doing so, the only phase is the uncancelled minus sign coming from the $n = 0$ terms -- this is just the minus sign from the classical calculation. In going to the last line we used the infinite products
\be
\prod_{n=1}^\infty c \to \frac{1}{\sqrt{c}}, \hspace{20pt} \prod_{n} n\to \sqrt{2\pi}, \hspace{20pt} \prod_{n=1}^\infty (1+\frac{x^2}{\pi^2n^2}) = \frac{\sinh(x)}{x}.
\ee
The final answer matches the negative term in (\ref{finite2}). 

\subsubsection*{Hilbert space interpretation for the maximal giant's thimble}

From the boundary perspective, the negative contribution $-q^{N+1}/(1-q)$ comes from finite $N$ trace relations. In \cite{Lee:2023iil}, this term and its generalizations were interpreted in the bulk theory as coming from negative norm (ghost) states of the giant graviton.\footnote{Candidate wave functions for these states were written in (2.30) of \cite{Lee:2023iil}, but we believe that those states actually correspond to the physical states with $R \le N$, see (\ref{physicalStates}) where $(r e^{\i\phi})_{\text{here}} = \phi_{\text{there}}$ and $(r e^{-\i\phi})_{\text{here}} = \bar\phi_{\text{there}}$.} We propose that these ghost states are the Hilbert space representation of the path integral thimble attached to the maximal giant graviton. 

Let's try to make this a bit more concrete. In our analysis of the BPS partition function, we started with a four-dimensional phase space $X_1,X_2,P_1,P_2$, but the $(X_2,P_2)$ oscillator got projected into its ground state $\langle X_2|0\rangle$. The remaining oscillator $(X_1,P_1)$ was integrated over a path integral thimble that quantizes the {\it imaginary continuation} of its phase space. The corresponding Hilbert space is the space of wave functions that are square-integrable for {\it imaginary} $X_1$. Within this space, the eigenstates of $R$ are
\be\label{states}
\langle X_1,X_2| N+1+n\rangle = \langle \i X_1|n\rangle\cdot \langle X_2|0\rangle = \texttt{Hermite}_n(\i X_1) \, e^{+X_1^2/2}\cdot e^{-X_2^2/2}.
\ee
Here, $\langle X|m\rangle$ means the ordinary harmonic oscillator wave function. Using (\ref{Hform}) and (\ref{Rform}) one can check that these states have $H-R = 1$ and $R = N+1+n$.

We claim that these states correspond to specific trace relations (columns with $N+1+n$ boxes) in the same way that the physical states of the giant graviton correspond to columns with $N$ or fewer boxes.

\section{A model for the half-BPS field space}\label{discussion}

Above, we studied the path integral thimble for a single maximal giant graviton. The half-BPS giant graviton expansion \cite{Gaiotto:2021xce} contains an infinite sequence of terms
\be \label{halfBPSgge}
\frac{1}{\prod_{n=0}^N (1-q^n)} = \frac{1}{\prod_{n=0}^\infty (1-q^n)} \sum_{k=0}^\infty (-1)^k q^{k N}\frac{q^{k(k+1)/2}}{\prod_{\ell=1}^k (1- q^\ell)}.
\ee
where the $k$-th term corresponds to the half-BPS spectrum of $k$-coincident giants. One could hope for a bulk description where the classical configurations with different numbers of giant graviton branes (including zero) are interpolated into each other, and the decomposition of the bulk path integral into thimbles could be made precise. We do not know of such a formalism, but we will discuss a model that comes somewhat close.\footnote{See section 8.1 \cite{Biswas:2006tj} for some relevant comments.}

In \cite{Biswas:2006tj,Beasley:2002xv}, it was argued that the phase space of D3 giant gravitons is given by $\CP^\infty$ with a symplectic form that is in the same cohomology class as
\be \label{CPsymform}
\Omega = (2 \pi N) \omega_{\rm FS},
\ee
where $\omega_{\rm FS}$ is the Fubini-Study form on $\CP^\infty$.\footnote{We take the Fubini-Study form $\omega_{\rm FS}$ on $\CP^\infty$ to be the $m \to \infty$ limit of those $\omega_{\rm FS}^{(m)}$ on $\CP^m$ satifying
\be
\int_{\CP^m} ( \omega_{\rm FS}^{(m)} )^m = 1.
\ee} In the half-BPS context, the idea is roughly that the coordinates on $\CP^\infty$ are coefficients parametrizing an arbitrary (anti-)holomorphic surface in one complex variable. It was shown there that an exact quantization of this phase space at finite $N$ yields the half-BPS Hilbert space $\cH_{N}$ and partition function
\be
\Tr_{\cH_{N}} q^{R} \, = \, \frac{1}{\prod_{n=1}^N (1 - q^n)}.
\ee

We are interested in the path integral formulation of the above statement. Let us observe that the above phase space is what one finds from considering a charged particle on the complex projective space $\CP^\infty$ with $N$ magnetic monopoles placed at the origin of the embedding space. By $N$ monopoles, we mean that the integral of the field strength across any $\CP^1$ inside $\CP^\infty$ is quantized in units of $N$ as
\be \label{dAquantization}
\int_{\CP^1} dA = 2\pi N.
\ee
From this, we may identify the field strength $\Omega = dA$ of the gauge field $A$ sourced by $N$ magnetic monopoles with the symplectic form \eqref{CPsymform}. The phase space path integral quantizing $\CP^\infty$ with symplectic form $\Omega$ and zero Hamiltonian is equivalent to the path integral of a charged particle on $\CP^\infty$ coupled only to the gauge field $A$ sourced by $N$ monopoles:
\be \label{modelpathint}
\Tr_{\cH_N} 1 =  \lim_{\mu\to 0} \ \int_{L\CP^\infty} Dz^a D\bar{z}^{\bar{a}} \, \sqrt{\det \Omega} \, \exp \left(-\mu \int_0^\beta \d \tau \, g_{a \bar{b}}\dot{z}^a \dot{\bar{z}}^{\bar{b}} + i \oint A \right).
\ee
The limit $\mu \to 0$ is equivalent to the low energy limit $\beta \to \infty$. We will naively set $\mu$ to zero for the moment but return to it later when we need to consider the limit more carefully.\footnote{The path integral with $\mu=0$ computes the Dirac index for the $\cO(N)$ line bundle over $\CP^m$, while the path integral in the $\mu \to 0$ limit computes its Dolbeault index. In our paper, we provide a derivation for the latter that uses $\mu$ as a regulator. But alternatively, the Dolbeault index can be computed directly from the path integral with $\mu=0$ by using that, on K\"{a}hler manifolds, the Dirac operator $\slashed{\pa}$ is the same as the Dolbeault operator $\bar{\pa}$ twisted by the square root of the canonical bundle $K$ \cite{Pestun:2016qko}: $\slashed{\pa} = \bar{\pa} \otimes K^{1/2}$. It holds that $K = \cO(-m-1)$ on $\CP^m$. Therefore, the Dolbeault index for the $\cO(N)$ bundle over $\CP^m$ can be computed by using a shifted value of $N \to N + \frac{m+1}{2}$ in the path integral with $\mu=0$.} The defining path integral contour is over the loop space $L\CP^\infty$, i.e. the space of maps of $S^1$ into $\CP^\infty$. The symplectic measure on $L\CP^\infty$ is induced from that on $\CP^\infty$ at each time $t$. It will be useful express \eqref{modelpathint} in local complex coordinates on $\CP^\infty$ as
\be \label{modelpathintsusy}
\int Dz^a D\bar{z}^{\bar{a}} D\psi^a D\bar{\psi}^{\bar{a}} \, \exp \left( - \int_0^\beta \d \tau \left[ -i (A_{a} \dot{z}^{a} + \bar{A}_{\bar{b}} \dot{\bar{z}}^{\bar{b}}) + i \Omega_{a \bar{b}} \psi^a \bar{\psi}^{\bar{b}} \right] \right)
\ee
While the original path integral \eqref{modelpathint} was that of a purely-bosonic system, the anti-commuting fields $\psi^a(\tau), \bar{\psi}^{\bar{b}}(\tau)$ arose due to a non-trivial symplectic form $\Omega$ in the field space $L\CP^\infty$. It will turn out that \eqref{modelpathintsusy} admits a ``hidden'' supersymmetry relating the bosons and anti-commuting fields. This property will allow us compute its partition function exactly as a sum over saddles and their one-loop fluctuations.

To be more explicit, it is helpful to recall some facts about complex projective spaces $\CP^m$. For the moment, we  take $m$ to be finite for regularization. $\CP^m$ is defined as the manifold obtained under the identification of homogeneous coordinates
\be
( w_0 : w_1 :  \cdots : w_m) \ \sim \ \lambda ( w_0 : w_1 :  \cdots : w_m)
\ee
for $\lambda$ a complex number. The metric of $\CP^m$ is invariant under $U(1)^{m+1}$ isometries that rotate each homogeneous coordinate $w_A$ independently as $w_A \to e^{i \alpha_A} w_A$. There are $m+1$ isolated fixed points under these isometries, with the $k$-th fixed point located at $w_k = 1$ and $w_{A \neq k} = 0$ where $k=0,1,2, \cdots, m$. Let us parametrize the patch $U_k$ centered at the $k$-th fixed point in terms of local coordinates
\be \label{eq: local coordinates CP}
z_1^{(k)} = \frac{w_0}{w_k}, \ z_2^{(k)} = \frac{w_1}{w_k}, \ \cdots \ , \ z_k^{(k)} = \frac{w_{k-1}}{w_k}, \ z_{k+1}^{(k)} = \frac{w_{k+1}}{w_k}, \ \cdots, \ z_m^{(k)} = \frac{w_m}{w_k}.
\ee
We will suppress labels indicating the patch $U_k$ for the local coordinates $z_a \equiv z_a^{(k)}$ when the patch is clear from context. In these coordinates, the Fubini-Study metric is
\ie
g_{a \bar{b}} &= \frac{1}{1 + z \cdot \bar{z}} \left[ \delta_{a \bar{b}} - \frac{\bar{z}^{\bar{a}} z^{b}}{(1 + z \cdot \bar{z})}  \right]
\fe
where $z \cdot \bar{z} \equiv \delta_{a \bar{b}} z^{a} \bar{z}^{\bar{b}}$, and the associated form is $\omega_{\rm FS} = \frac{i}{2 \pi} \sum_{a,b} g_{a \bar{b}} dz^a \wedge d\bar{z}^{\bar{b}}$.

We can write down the gauge field $A^{(k)}$ sourced by $N$ monopoles on $U_k$ by requiring that it is locally a primitive $\Omega = dA^{(k)}$ of the symplectic form $\Omega = (2 \pi N) \omega_{\rm FS}$. This is so that the integral of the pullback of the field strength over any $\CP^1 \subset \CP^m$ satisfies \eqref{dAquantization}. In a gauge that is non-singular at the origin of $U_k$, it is
\be
A^{(k)} = \frac{i N}{2} \frac{1}{1 + z \cdot \bar{z}} \left( z \cdot d\bar{z} - \bar{z} \cdot dz \right).
\ee
It is straightforward to compute how gauge fields on the different patches are related. Gauge fields on the intersection $U_k \cap U_{k+1}$ are related as
\be \label{Atransition}
A^{(k)} - A^{(k+1)} = \frac{i N}{2} \left( \frac{dz_{k+1}^{(k+1)}}{z_{k+1}^{(k+1)}} - \frac{d\bar{z}_{k+1}^{(k+1)}}{\bar{z}_{k+1}^{(k+1)}} \right).
\ee
The difference between those on any two patches $U_k, U_\ell$ may be found on $U_k \cap U_\ell$ by subsequent application of \eqref{Atransition}. The integral of this difference
\be
i \oint_\gamma ( A^{(k)} - A^{(k+1)} ) = 2 \pi i N,
\ee
along a topologically non-trivial closed path $\gamma$ wound once in $U_k \cap U_{k+1}$, restricts the monopole charge $N$ to be an integer for the path integral to be well-defined under transition maps.

How are the coordinates on $\CP^m$ charged under the $U(1)_R$ R-symmetry relevant in the half-BPS sector? As mentioned earlier, the homogeneous coordinates $w_A$ are coefficients that parametrize an arbitrary (anti-)holomorphic surface $P(\bar{Z}) = \sum_{n=0}^m w_n \bar{Z}^n = 0$ in one complex variable $\bar{Z}$. The variable $\bar{Z}$ in this context is a coordinate on a complex plane $\bC$ inside $\bC^3$, and the intersection of $P(\bar{Z}) = 0$ with the unit $S^5$ in $\bC^3$ is viewed as the worldvolume of a giant graviton \cite{Biswas:2006tj,Beasley:2002xv,Mikhailov:2000ya}.\footnote{Our conventions differ slightly from that in \cite{Biswas:2006tj}, stemming from the fact that the Fubini-Study form $\omega_{\rm FS}$ used there actually differs by an overall sign from the standard one used here. One downstream consequence is that $w_A$ here must be coefficients of an anti-holomorphic surface, rather than a holomorphic one, for consistency with the path integral computation. The R-charge operator $L$ used in \cite{Biswas:2006tj} indeed assigns positive R-charges to $w_A$, consistent with charge assignments here for an anti-holomorphic surface.} Therefore if $\bar{Z}$ has R-charge $R = -1$, the homogeneous coordinate $w_A$ on the phase space possesses R-charge $R = A$. It follows that the local coordinates $z_a^{(k)}$ on the patch $U_k$ are charged under the insertion of the operator $q^R$ in the trace over the Hilbert space as
\be \label{Rassignment}
( q^{-k} z_1^{(k)}, \ q^{-k+1} z_2^{(k)}, \ \cdots , \ q^{-2} z_{k-1}^{(k)}, \ q^{-1} z_k^{(k)}, \ q^1 z_{k+1}^{(k)}, \ q^2 z_{k+2}^{(k)}, \ \cdots),
\ee
while $\bar{z}_{\bar{b}}^{(k)}$ carry opposite charges. The anti-commuting fields $\psi_a^{(k)},\bar{\psi}_{\bar{b}}^{(k)}$ are charged in the same manner as the partner bosons.

Let us return to the path integral \eqref{modelpathint} of a charged particle on $\CP^m$. The Euclidean path integral on the patch $U_k$ is
\be \label{unrefinedeucl}
\Tr_{\cH_N} 1 = \int Dz^a D\bar{z}^{\bar{a}} D\psi^a D\bar{\psi}^{\bar{a}} \, \exp \left( - N \int_0^\beta \d \tau \left[ \frac{1}{2} \frac{z \cdot \dot{\bar{z}} - \bar{z} \cdot \dot{z} }{1+ z\cdot \bar{z}} - \frac{1}{1 + z \cdot \bar{z}} \left( \psi \cdot \bar{\psi} - \frac{(\bar{z} \cdot \psi)(z \cdot \bar{\psi})}{1 + z \cdot \bar{z}} \right) \right] \right).
\ee
We are interested in computing the quantity $\Tr_{\cH_N} q^R$, so let us introduce a chemical potential $\log \frac{1}{q}$ for the R-charge. Then the action on patch $U_k$ becomes, after rescaling time to set $\beta = 1$,
\ie \label{refinedaction}
S^{(k)} = S_0^{(k)} + N \int_0^1 &\d \tau \Bigg[ \frac{1}{2} \frac{z \cdot \dot{\bar{z}} - \bar{z} \cdot \dot{z} }{1+ z\cdot \bar{z}} - \frac{1}{1 + z \cdot \bar{z}} \left( \psi \cdot \bar{\psi} - \frac{(\bar{z} \cdot \psi)(z \cdot \bar{\psi})}{1 + z \cdot \bar{z}} \right) \\
&\hspace{-10pt}+ \frac{1}{1+ z \cdot \bar{z}} \left( - \sum_{a=1}^k (k-a+1) \log \tfrac{1}{q} \, z^a \bar{z}^{\bar{a}} + \sum_{a=k+1}^m (a-k) \log \tfrac{1}{q} \, z^a \bar{z}^{\bar{a}} \right) \Bigg].
\fe
Here $S_0^{(k)}$ is a constant that we will determine shortly. This action has $m+1$ saddle points, corresponding to the $m+1$ fixed points of the $U(1)^{m+1}$ isometries.\footnote{Classical solutions in polar coordinates $z^a = r_a e^{i \phi_a}$ and $\bar{z}^a = r_a e^{-i \phi_a}$ are paths of constant $r_a$ and of constant imaginary angular velocity $\dot{\phi}_a(\tau)$ (when $\log \frac{1}{q}$ is assumed to be real) whose value is proportional to its mass \eqref{refinedaction}, for all $a$. However, the only solutions consistent with the periodic boundary conditions of our Euclidean path integral are the constant maps from $S^1$ to the origin $z_a^{(k)}(\tau) = \bar{z}_{\bar{a}}^{(k)}(\tau) = 0$ of each patch $U_k$. Thus we may restrict to considering path integrals over the Lefschetz thimbles associated to these maps only.} From the perspective of each patch, there is one obvious saddle point at $z = 0$, and then there are less-obvious saddle points where one of the $z^a$ coordinates goes to infinity. The saddle point with $z^a =\infty$ within the $U_{k = 0}$ patch maps to the saddle with $z = 0$ in the patch $U_{k = a}$. Consistency of the action between patches then requires
\be
S_0^{(k)} \ = \ k N \log \tfrac{1}{q},
\ee
up to an overall shift that we will use this to absorb the $\mu$-dependent ground state energy.

This path integral is one-loop exact by the Duistermaat-Heckman theorem \cite{Duistermaat:1982vw} \footnote{See also \cite{Szabo:1996md} and references therein and, e.g. \cite{Stanford:2017thb}, for a more recent application.} because we are integrating over a symplectic manifold the exponential of an action that generates a $U(1)$ symplectomorphism, generated by the vector field 
\ie
V^{(k)} = \frac{d}{d\tau} + \log \tfrac{1}{q} \Bigg[ \sum_{a=1}^k &(k-a+1) \left( z^a(\tau) \frac{\pa}{\pa z^a(\tau)} - \bar{z}^{\bar{a}}(\tau) \frac{\pa}{\pa \bar{z}^{\bar{a}}(\tau)} \right) \\
&- \sum_{b=k+1}^m (b-k) \left( z^b(\tau) \frac{\pa}{\pa z^b(\tau)} - \bar{z}^{\bar{b}}(\tau) \frac{\pa}{\pa \bar{z}^{\bar{b}}(\tau)} \right) \Bigg].
\fe
Concretely, the action \eqref{refinedaction} has a supersymmetry
\be
\delta z^a = \psi^a, \quad \delta \bar{z}^{\bar{a}} = \bar{\psi}^{\bar{a}}, \quad \delta \psi^a = - \dot{z}^a - (k-a+1) \log \tfrac{1}{q} \, z^a, \quad \delta \bar{\psi}^{\bar{a}} = - \dot{\bar{z}}^{\bar{a}} + (k-a+1) \log \tfrac{1}{q} \, \bar{z}^{\bar{a}}
\ee
for $0 \leq a \leq k$ and
\be
\delta z^a = \psi^a, \quad \delta \bar{z}^{\bar{a}} = \bar{\psi}^{\bar{a}}, \quad \delta \psi^a = - \dot{z}^a + (a-k) \log \tfrac{1}{q} \, z^a, \quad \delta \bar{\psi}^{\bar{a}} = - \dot{\bar{z}}^{\bar{a}} - (a-k) \log \tfrac{1}{q} \, \bar{z}^{\bar{a}}
\ee
for $a>k$. 
Therefore, we can evaluate the path integral for $\Tr_{\cH_N} q^R$ on $\CP^m$ as a sum over one-loop determinants around the $m+1$ saddle points:
\be \label{saddlesum}
\Tr_{\cH_N^{(m)}} \, q^R  \ = \ \sum_{k=0}^m e^{-S_0^{(k)}} \int Dz^a D\bar{z}^{\bar{a}} e^{- S_{\rm quad}^{(k)}(z,\bar{z})}
\ee
with (rescaling the fluctuation fields as $(z,\bar{z}) \to \frac{1}{\sqrt{N}}(z,\bar{z})$.)
\be \label{quadaction}
S_{\rm quad}^{(k)} = \int_0^1 \d \tau \left[ \frac{1}{2} (z \cdot \dot{\bar{z}} - \bar{z} \cdot \dot{z}) - \sum_{a=1}^k (k-a+1) \log \tfrac{1}{q} \, z^a \bar{z}^{\bar{a}} + \sum_{a=k+1}^m (a-k) \log \tfrac{1}{q} \, z^a \bar{z}^{\bar{a}} \right].
\ee

In the $m \to \infty$ limit, $S_{\rm quad}^{(k)}$ is a functional on the field space with an infinite number of stable constant modes and with a finite number $2k$ of constant modes that are unstable. To evaluate the one-loop determinant, we introduce back the regulator term \eqref{modelpathint} near the $k$-th saddle
\be
S_{\rm reg}^{(k)} = \mu \int_0^1 \d \tau \ D_\tau z \cdot D_\tau \bar{z}
\ee
and consider the $\mu \to 0$ limit. We included the effect of the chemical potential by defining $D_\tau z^a = (\pa_\tau - \sigma_a) z^a$ and $D_\tau \bar{z}^{\bar{a}} = (\pa_\tau + \sigma_a) \bar{z}^{\bar{a}}$ where
\be
\sigma_a \ = \ \begin{cases} \ -(k-a+1) \log \tfrac{1}{q} & \quad 1 \leq a \leq k \\
\ +(a-k) \log \tfrac{1}{q} & \quad a > k \end{cases}.
\ee
The path integral for the $a$-th fluctuation fields in $S_{\rm reg}^{(k)} + S_{\rm quad}^{(k)}$ evaluates to
\ie
\int Dz^a D\bar{z}^{\bar{a}} \ \exp &\left( - \int_0^1 \d \tau \left[ \mu \, D_\tau z^a D_\tau \bar{z}^{\bar{a}} + \frac{1}{2}(z^a D_\tau \bar{z}^{\bar{a}} - \bar{z}^{\bar{a}} D_\tau z^{a} ) \right] \right) \\
&\xrightarrow{\mu \to 0} \ \ \begin{dcases*} \ \frac{- e^{\sigma_a}}{1 - e^{\sigma_a}} e^{-\frac{1}{2\mu}} & $\quad 1 \leq a \leq k$ \\
\ \frac{1}{1 - e^{-\sigma_a}} e^{-\frac{1}{2\mu}} & $\quad a > k$ \end{dcases*}
\fe
and we absorb the $-\frac{1}{2\mu}$ term by shifting the zero-point energy. So the stable directions contribute
\be
\frac{1}{\prod_{n=1}^{m-k} \left( 1 - q^{n} \right)}
\ee
while the unstable directions, with their rotated steepest-descent contours, contribute
\be
\frac{(-1)^k q^{\frac{k(k+1)}{2}}}{\prod_{\ell=1}^k \left( 1 - q^\ell \right)}
\ee
with regularization as in Section \ref{quantum}. The signs $(-1)^k$ accompanying the saddles result from decomposing the defining integration contour of the path integral into steepest-descent contours which, locally around the $k$-th saddle, involve $2k$ constant bosonic modes that are Wick-rotated relative to the defining contour.

The above property is reminiscent of what we found in our analysis of the thimble associated with a maximal giant. In fact, let us write the action for the fluctuation fields $z^1,\bar{z}^{1}$ in $S_{\rm quad}^{(k=1)}$, which are the only fields near the $k=1$ saddle whose steepest-descent contours are subject to rotation:
\ie
S_{\rm quad}^{(k=1)}(z^1,\bar{z}^1) &= \int_0^1 \d \tau \left[ \frac{1}{2} (z^1 \dot{\bar{z}}^{1} - \bar{z}^{1} \dot{z}^1) - \log \tfrac{1}{q} \, z^1 \bar{z}^{1} \right] \\
&= \int_0^1 \d \tau \left[ -i P \dot{X} - \frac{\log \frac{1}{q} }{2} (X^2 + P^2) \right],
\fe
where we defined $z^1 = \frac{1}{\sqrt{2}}(P + i X)$ and $\bar{z}^1 = \frac{1}{\sqrt{2}}(P - i X)$. The action for unstable fluctuation fields $z^1,\bar{z}^{1}$ turns out to be the phase space action \eqref{pathint7} relevant for half-BPS excitations of the maximal giant, expressed in the holomorphic representation. There is also a bulk interpretation for the stable contributions in our path integral model: their spectrum in the $m \to \infty$ limit coincides with the spectrum of half-BPS Kaluza-Klein modes.\footnote{Given the matching of spectra, it is natural to ask if the action for unstable directions at higher $k$ relates to that of half-BPS excitations on $k$-coincident maximal giants. However, the latter involves $U(k)$-gauged matrices so the connection is not obvious. An effective action for half-BPS modes on $k$ giants and its relation to the unstable directions near the $k$-th saddle would interesting to study, possibly along the lines of \cite{Gopakumar:2024jfq}.}

To summarize, the $k$-th saddle point of the path integral quantizing the BPS phase space $\CP^{\infty}$ with symplectic form $\Omega = (2 \pi N) \omega_{\rm FS}$ has the bulk interpretation of $AdS_5 \times S^5$ with $N$ units of five-form flux and with $k$-coincident maximal half-BPS giant gravitons. The stable fluctuations around the $k$-th saddle are interpreted as Kaluza-Klein excitations, while the unstable fluctuations (with $2k$ of its constant bosonic modes possessing contours that are rotated away from the defining contour) have the interpretion of open strings on $k$ giants. The defining contour of this Euclidean path integral, in which $N$ enters only as a coupling, is decomposed exactly into a sum over the $k=0,1,2,\cdots$ saddles and their ``one-loop'' thimbles.

Let us finally evaluate the path integral quantizing the BPS phase space as a sum over the fixed points of $\CP^m$:\footnote{Generalization of our path integral analysis to the $1/4$- and $1/8$-BPS chiral ring sectors considered in \cite{Biswas:2006tj} is straightforward. In the $1/8$-BPS case, the homogeneous coordinates $w_{(n_1,n_2,n_3)}$ of $\CP^\infty$ parametrize the surface $\sum_{n_1,n_2,n_3} w_{(n_1,n_2,n_3)} \bar{Z}_1^{n_1} \bar{Z}_2^{n_2} \bar{Z}_3^{n_3} = 0$ with three independent $U(1)_R$ charges.}
\be
\Tr_{\cH_N^{(m)}} \, q^R  \ = \ \sum_{k=0}^m \frac{(-1)^k q^{k N+ \frac{k(k+1)}{2}}}{\prod_{n=1}^{m-k} \left( 1 - q^{n} \right) \, \prod_{\ell=1}^k \left( 1 - q^{\ell} \right)}.
\ee
In the $m \to \infty$ limit, it becomes the expansion \eqref{halfBPSgge} of the half-BPS partition function in terms of imaginary thimbles associated with maximal giants:
\be
\Tr_{\cH_N} q^R \ = \ \frac{1}{\prod_{n=0}^\infty (1-q^n)} \sum_{k=0}^\infty (-1)^k q^{k N}\frac{q^{k(k+1)/2}}{\prod_{\ell=1}^k (1- q^\ell)}
\ee
while the $m=1$ case recovers the baby example \eqref{finite} in the introduction:
\be
\Tr_{\cH_N^{(1)}} \, q^R \ = \ \frac{1}{1 - q} - \frac{q^{N+1}}{1-q}.
\ee

\section*{Acknowledgements} 

We are grateful to Matteo Beccaria, Davide Gaiotto, Shota Komatsu, Raghu Mahajan, Edward Mazenc, Sameer Murthy, and Zhenbin Yang for discussions. This work was supported in part by DOE grant DE-SC0021085, by the Sloan Foundation, and by a grant from the Simons foundation (926198, DS). The work of the group at ETH is supported by the Simons Foundation grant 994306 (Simons Collaboration on Confinement and QCD Strings) and by the NCCR SwissMAP funded by the Swiss National Science Foundation. This work was also supported in part by the Simons Foundation grant 994308.

\appendix

\section{Unstable saddle point on the defining contour}\label{app:unstable}
Consider an integral 
\be
Z(\beta) = \int_{-\infty}^\infty \d x \ e^{-\beta I(x)}
\ee
where the potential function has the following form:
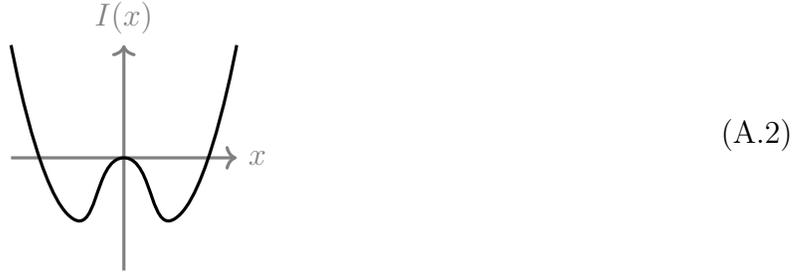
\be
\begin{tikzpicture}[scale=0.75, rotate=0, baseline={([yshift=-0.15cm]current bounding box.center)}]
\draw[very thick, ->, gray] (-2,0) -- (2,0) node[right] {$x$};
\draw[very thick, ->, gray] (0,-2) -- (0,2) node[above] {$I(x)$};
\draw[very thick] plot [smooth, tension=1] coordinates { (-2,2) (-1,-1) (0,0) (1,-1) (2,2)};
\end{tikzpicture}
\ee
For large real $\beta$, the dominant contribution comes from the stable saddle points at the two minima of the potential. There is also an unstable saddle point at the origin. Does it contribute to the integral?


The coefficient with which a saddle point contributes to an integral is called the Stokes multiplier. In the generic case where no saddle points have the same imaginary part of their action, the Stokes multiplier of a given saddle point is the signed intersection number between the {\it upward} flow from that saddle and the defining contour (see section 3 of \cite{Witten:2010cx}). For our integral, the upward flow from the origin coincides with the defining contour and the intersection number is not defined. One way to proceed is to jiggle the integral slightly by adding a small imaginary part to $\beta$. Then the upward flow manifold starting from the origin will pass through the origin with an angle that is rotated slightly away from the real axis (clockwise or counterclockwise depending on the sign of the imaginary part of $\beta$) -- the Stokes multiplier is $\text{sgn}(\text{Im}(\beta))$ and the one-loop contribution of the saddle is
\be\label{contri}
\text{sgn}(\text{Im}(\beta)) \int_{-\i\infty}^{\i\infty} \d x e^{-\beta I(0) - \frac{1}{2}x^2 I''(0)}.
\ee
The case of real $\beta$ is right on a Stokes line, where the multiplier is somewhat ambiguous. This ambiguity corresponds to the ambiguity in the definition of the dominant Lefschetz thimbles associated to the minima of $I(x)$. For real $\beta$, these downward flow thimbles hit our unstable saddle. A natural choice is to define the thimble for the dominant saddle to be the average of the well-defined thimbles on the two sides of the Stokes line. This means that we should take the Stokes multiplier for the unstable saddle to be the average of its value for positive and negative imaginary parts of $\beta$. In our case this gives zero -- the unstable saddle does not contribute.

This conclusion seems intuitive because for real $\beta$ the contribution of the unstable saddle point would be imaginary, while the entire integral is real. So unless we define the contributions of the dominant saddles with a prescription that makes them complex, the unstable saddle must not contribute.

Let's now consider an analogous problem in $d$ dimensions, where we have an integral over $\{x_1,\dots,x_d\}$ where the defining contour is the real $x$ plane
\be
Z(\beta) = \int \d^d x \ e^{-\beta I(x)}
\ee
and $I(x)$ has an unstable saddle point near the origin (for simplicity we can consider the case where it is rotationally symmetric):
\be
I(x) = -\frac{1}{2}x\cdot x + \dots.
\ee
We can follow the same procedure of adding a small imaginary part to $\beta$. The Stokes multiplier now be $[\text{sgn}(\text{Im}(\beta))]^d$. If $d$ is odd, then it flips sign depending on the imaginary part of $\beta$, but if $d$ is even it is always one. So for odd $d$ the unstable saddle does not contribute and for even $d$ it contributes with Stokes multiplier one. Including the one-loop determinant and the $d$ factors of $\i$ from the rotation of the contour, we find the contribution (for real $\beta$)
\be\label{contribution}
\text{odd }d: \hspace{5pt}0, \hspace{40pt} \text{even }d: \hspace{5pt}\left(-\frac{2\pi}{\beta}\right)^{\frac{d}{2}}.
\ee
Note that there is no contradiction with reality because the contribution of the unstable saddle point is proportional to $i^d$ which is real for even $d$.

Let's now discuss examples of this. Consider the integral
\be
Z_d(\beta) = \int_{S^d} e^{\beta (1+\cos\theta)} = \frac{(2\pi)^{\frac{d+1}{2}}}{\beta^{\frac{d-1}{2}}}e^\beta I_{\frac{d-1}{2}}(\beta).
\ee
There is a dominant saddle at $\theta = 0$ and a subdominant unstable saddle at $\theta = \pi$. For even $d$ the Bessel function simplifies and it is easy to see that for large $\beta$ the unstable saddle contributes with coefficient (\ref{contribution}). For example, for $d = 2$ and $d = 4$, the integral is
\begin{align}
Z_2(\beta) &= \frac{2\pi}{\beta}e^{2\beta} \boxed{- \frac{2\pi}{\beta}}\\
Z_4(\beta) &= \frac{4\pi^2}{\beta^2}e^{2\beta}\left(1 - \frac{1}{\beta}\right) +\boxed{ \frac{4\pi^2}{\beta^2}\left(1 + \frac{1}{\beta}\right)}.
\end{align}
The boxed terms are the contributions of the unstable saddle and their coefficients are compatible with (\ref{contribution}). These examples are particularly clear because the loop expansion truncates, so the exact answer can be unambiguously separated into contributions from the two saddle points. For a trickier example, consider e.g.
\be
\int_{\mathbb{R}^2}e^{\beta(\frac{x\cdot x}{2} - \frac{(x\cdot x)^3}{6})} = \pi \int_0^\infty \d y \ e^{\beta(y/2 - y^3/6)}.
\ee
An experimental way to see the contribution of the unstable saddle point is to do optimal truncation of the asymptotic series around the dominant saddle point at $y = 1$ \cite{berry1989uniform}. This means cutting off the series at the smallest term. For $\beta = 20$, the smallest term is the 15th term in the perturbative series, which contributes $0.000077\dots$. Summing the dominant series up to this point gives the estimate $445.4566\dots$ for the integral, whereas the exact answer is $445.3540\dots$. The leading (one-loop) contribution of the unstable saddle is $-0.1$, which significantly improves the result of the optimally truncated dominant series. The two-loop contribution around the unstable saddle gives $-0.002$, which improves the result further.

For odd $d$, the contribution of the unstable saddle is imaginary, and it is clear that it will never improve the approximation of an optimally truncated series around the dominant saddle, which will be real for real $\beta$.

\section{A model integral}
It is interesting to compare our discussion of the maximal giant to the famous integral
\be\label{int}
\int_{S^2}e^{\mu \cos\theta } = \frac{2\pi}{\mu}\left(e^{\mu} - e^{-\mu}\right).
\ee
There are at least three ways of thinking about this integral, and they each offer a different interpretation of the minus sign multiplying the second term.
\begin{enumerate}
\item After integrating over $\phi$ and changing variables from $\theta$ to $\cos(\theta)$ one gets
\be\label{method1}
\int_0^{2\pi} \d\phi\int_{0}^{\pi}\d\theta\sin(\theta) e^{\mu \cos\theta} = 2\pi\int_{-1}^1\d(\cos\theta)e^{\mu\cos\theta}.
\ee
The two terms in (\ref{int}) are endpoint contributions from the north and south poles. The $e^{-\mu}$ term has a minus sign because it is a lower endpoint. 

\item From a smooth two-dimensional perspective there are no endpoints, but there are saddle points at the north and south poles. The north pole is as stable saddle that gives the positive term, and the south pole is an unstable saddle. The minus sign $-1 = \i^2$ comes from Wick rotating two coordinates to follow the Lefschetz thimble attached to this saddle. 

\item One can write the $\sin(\theta)$ measure as an integral over fermionic variables $\d\theta$ and $\d\phi$:
\be
\int \d\theta\d\phi \d(\d\theta)\d(\d\phi) \exp\left[\mu\cos(\theta) + \sin(\theta)\d\phi\d\theta\right].
\ee
This integral has a supersymmetry under
\be
Q\theta = \d\theta, \hspace{20pt} Q\phi = \d\phi, \hspace{20pt} Q(\d\theta) = 0, \hspace{20pt} Q(\d\phi) = \mu.
\ee
Adding a term $Q(\sin^2(\theta)\d\phi)$ to the action will not change the answer for the integral, but it does change the integrand. After adding this term, we have
\be
\int \d\theta\d\phi \d(\d\theta)\d(\d\phi) \exp\left[\mu\cos(\theta) -\lambda\sin^2(\theta)+ \sin(\theta)\big(1 + 2\lambda\cos(\theta)\big)\d\phi\d\theta\right].
\ee
The north and south poles are still saddle points of this action. But if we make $\lambda$ sufficiently large, they are both stable. The minus sign comes from the fermionic integral. 
\end{enumerate}

Coming back to the giant graviton, we regard this integral as a toy model for the sum over BPS states corresponding to a single column. Method 1) is analogous to the straightforward sum over physical BPS states, with a cutoff corresponding to the maximal giant. Method 2) is analogous to the perspective discussed in this paper, where the maximal giant is an unstable saddle point. Method 3) may be related to the perspective in \cite{Eleftheriou:2023jxr}.\footnote{However, one would need to add a localizing term that makes the maximal giant a stable saddle point. \cite{Eleftheriou:2023jxr} studies the original giant graviton action for which the physical states near the giant graviton (\ref{physicalStates}) should be bosonic and should have $R\le N$. We believe the formula $R = -\frac{\i}{L}\partial_{\varphi}$ that follows (3.8) of \cite{Eleftheriou:2023jxr} should read $R = +\frac{\i}{L}\partial_{\varphi}$, because from (2.21) of that paper one can see that $R$ should be (up to a constant shift) the momentum conjugate to $\phi$, or minus the momentum conjugate to $\varphi$.}

This model integral gives some perspective on the ``physicality'' of the states (\ref{states}) associated to the maximal giant's thimble. Let's use the form
\be
2\pi\int_0^\pi \d\theta\sin(\theta)e^{\mu\cos\theta}.
\ee
If $\mu$ has a small positive imaginary part (see appendix \ref{app:unstable} for more on this) then the defining contour decomposes into two steepest descent contours (Lefschetz thimbles) for the stable and unstable saddle points, respectively:
\be \label{contourequation}
\begin{tikzpicture}[scale=0.75, rotate=0, baseline={([yshift=-0.15cm]current bounding box.center)}]
\draw[ ->, gray] (-1,0) -- (1.8,0);
\draw[->, gray] (0,-1.8) -- (0,1.8);
\draw[gray] (1.4,1.9) -- (1.4,1.4) -- (1.9,1.4);
\node[] at (1.6,1.67) {$\theta$};
\draw[very thick, ->] (0,0) -- (.85,0);
\draw[very thick] plot [smooth, tension=1] coordinates { (0,0) (1.5,0)};
\draw[very thick] (1.5,-.1) -- (1.5,.1) node[above] {$\pi$};
\draw[very thick] (0,-.1) -- (0,.1) node[above=5, left=.2] {$0$};
\end{tikzpicture} \hspace{20pt} = \hspace{20pt}\begin{tikzpicture}[scale=0.75, rotate=0, baseline={([yshift=-0.15cm]current bounding box.center)}]
\draw[ ->, gray] (-1,0) -- (1.8,0);
\draw[->, gray] (0,-1.8) -- (0,1.8);
\draw[very thick, ->] (.8,-.03) -- (.85,-.033);
\draw[very thick] plot [smooth, tension=.5] coordinates { (0,0) (1.3,-.2) (1.5,-1.8)};
\end{tikzpicture} \hspace{20pt}+ \hspace{20pt}\begin{tikzpicture}[scale=0.75, rotate=0, baseline={([yshift=-0.15cm]current bounding box.center)}]
\draw[ ->, gray] (-1,0) -- (1.8,0);
\draw[->, gray] (0,-1.8) -- (0,1.8);
\draw[very thick, ->] (1.5,-.9) -- (1.5,-.8);
\draw[very thick] plot [smooth, tension=1] coordinates { (1.5,-1.8) (1.5,0)};
\end{tikzpicture}
\ee
If the ``physical'' contribution is the defining contour, then the contribution of the unstable saddle point (analogous to the maximal giant's thimble) is {\it not} physical. Instead, it exists to correct a kind of overcounting mistake from the contour of the stable saddle point, just as trace relations fix an overcounting mistake in the $N=\infty$ spectrum of Kaluza-Klein modes.

\bibliography{references}

\bibliographystyle{utphys}

\end{document}